\pdfoutput=1
\documentclass[a4paper,fleqn,usenatbib]{mnras}

\usepackage{newtxtext,newtxmath}
\usepackage[T1]{fontenc}
\usepackage{ae,aecompl}

\newcommand\blfootnote[1]{%
  \begingroup
  \renewcommand\thefootnote{}\footnote{#1}%
  \addtocounter{footnote}{-1}%
  \endgroup
}

\usepackage{graphicx}	% Including figure files
\usepackage{amsmath}	% Advanced maths commands
\usepackage{amssymb}	% Extra maths symbols

\title[Astro-Classification with GRU Ensembles]{Astronomical Classification of Light Curves with an Ensemble of Gated Recurrent Units}

\author[S. Chaini and S. S. Kumar]{
Siddharth Chaini,$^{1}$\thanks{E-mail: siddharthc17@iiserb.ac.in}
Soumya S. Kumar,$^{1}$\thanks{E-mail: soumya17@iiserb.ac.in}
\\
$^{1}$Department of Physics, Indian Institute of Science Education and Research, Bhopal\\
}

\date{Accepted XXX. Received YYY; in original form ZZZ}

\pubyear{2020}

\begin{document}
\label{firstpage}
\pagerange{\pageref{firstpage}--\pageref{lastpage}}
\maketitle

\begin{abstract}
With an ever-increasing amount of astronomical data being collected, manual classification has become obsolete; and machine learning is the only way forward. Keeping this in mind, the Large Synoptic Survey Telescope (LSST) Team hosted the Photometric LSST Astronomical Time-Series Classification Challenge (PLAsTiCC) in 2018. The aim of this challenge was to develop models that accurately classify astronomical sources into different classes, scaling from a limited training set to a large test set. In this text, we report our results of experimenting with Bidirectional Gated Recurrent Unit (GRU) based deep learning models to deal with time series data of the PLAsTiCC dataset. We demonstrate that GRUs are indeed suitable to handle time series data. With minimum preprocessing and without augmentation, our stacked ensemble of GRU and Dense networks achieves an accuracy of 76.243\%. Data from astronomical surveys such as LSST will help researchers answer questions pertaining to dark matter, dark energy and the origins of the universe; accurate classification of astronomical sources is the first step towards achieving this. 

Our code is open-source and has been made available on GitHub here: \url{https://github.com/AKnightWing/Astronomical-Classification-PLASTICC}

\end{abstract}

\begin{keywords}
methods: data analysis -- techniques: photometric
\end{keywords}

%%%%%%%%%%%%%%%%%%%%%%%%%%%%%%%%%%%%%%%%%%%%%%%%%%

%%%%%%%%%%%%%%%%% BODY OF PAPER %%%%%%%%%%%%%%%%%%

\section{Introduction}

\blfootnote{Note: Any bugs can be reported on Github itself.}
Over the last decade, numerous large scale astronomical surveys have been conducted to systematically collect images of the night sky using various spectroscopic and photometric methods. These surveys have, in turn, led to the discovery of an unprecedented number of transients as well as variable astronomical objects. However, with the ever-increasing size of available data, these surveys have also brought to light the problem of astronomical classification for big data.

NASA's Kepler Space Telescope, designed to determine the occurrence rate of Earth-sized planets in temperate orbits around Sun-like stars, photometrically observed about 200,000 stars (\citet{Jenkins_2010}; \citet{Koch_2010}; \citet{Christiansen_2015}) and discovered thousands of transiting exoplanets (\citet{borucki2011characteristicsa}, \citet{borucki2011characteristicsb}; \citet{batalha2011kepler}; \citet{burke2014planetary}; \citet{rowe2014validation}). In the first stage of its operation, Sloan Digital Sky Survey (SDSS; \citet{Frieman_2007}) measured the spectra of more than 700,000 celestial objects, and the SDSS Supernova Survey measured light curves for a few hundred supernovae, of which spectroscopic confirmations for 500 SN Ia and about 80 core-collapsed supernovae were obtained. Between 2013 and 2019, the Dark Energy Survey (DES; \citet{darkenergysurvey}) recorded information from about 300 million galaxies. The Zwicky Transient Facility (ZTF; \citet{Kulkarni2016}), a collaboration project by Caltech and other notable institutions, had first light at Palomar Observatory in 2017. ZTF produces around one terabyte (TB) of raw image data and 4 TB of real-time data products each night (on an average uninterrupted observing night spanning roughly 8 hr 40 mins). Over the course of the nominal three-year survey, this would amount to more than 50 TB of light curve data, 60 TB of reference image products; with the total volume of data amounting to around 3.2 petabytes (PB) (assuming 260 good weather observing nights) \citep{Masci_2018}. ZFT will form the basis of even larger surveys such as the LSST which will build on ZFT's rapid scans of the sky. The enormous Vera C. Rubin Telescope and its ambitious Legacy Survey of Space and Time will usher in a new-age by generating more than 20 TB of data per night, with the final dataset expected to weigh in at 15 PB. During its ten-year survey duration, it is expected to observe $ 2 \times 10^{10}$ galaxies, $ 1.7 \times 10^{10} $ resolved stars and discover $10^7$  supernovae \citep{collaboration2009lsst}. 

These and other surveys have been generating data in various forms in very large quantities; and as technological progress is made in this field, the rate of data generation is only set to increase. Classification of celestial objects by astronomers depends heavily on obtaining their spectroscopic light curves, but extensive spectroscopic data is hard to come by due to the lack of availability of resources. This makes it even more crucial to identify objects of interest for spectroscopic follow-up. Traditionally, astronomers used to classify objects based on visual inspection, but this quickly became a bottleneck as the rate of data generation increased drastically. Numerous citizen-science projects were started to speed up this process by roping in amateur astronomers to help with classification. But the speed of human classification was soon subjugated by the sheer amount of raw data available. The situation called for an intervention utilising computational techniques to automate the process in its initial stages to speed up the rate of object classification as well as to ensure optimal allocation of spectroscopic resources to objects of interest. 

The nature of the problem proved to be most suitable for the application of machine learning techniques. One of the first examples of machine learning techniques in astronomy dates back to \citet{bailey2007find}. They reported on object classification for supernovae using the Supernovae Factory data with synthetic supernovae as training data. Since photometric data was readily available, the focus centred on using photometric data for classification. The Supernova Photometric Classification Challenge by DES (SNPhotCC; \citet{kessler2010supernova}) was the first of its kind with a spectroscopically confirmed training set of 1,103 objects and a test set of 20,216 objects without spectroscopic confirmation. Participants were challenged to develop classifiers that could use the known labels of the training set to infer the types of objects in the test set. SNPhotCC paved the way for people from a non-astronomy background to approach the problem of astronomical classification, allowing people to come up with fresh, innovative approaches to this problem. 

In astronomy, Cepheid variables and Type Ia Supernovae (SNIa) are used as \emph{standard candles}. Standard candles are sources that have a known luminosity and are used to measure distances. It was measurements with SNIa that led to the discovery of the accelerating expansion of the universe, and thus the discovery of dark energy. Photometry of Type Ia supernovae as a function of redshift serves as a powerful probe to aid studies and arrive at important cosmological constraints. Pure samples of Type Ia supernovae will prove to be crucial to arrive at statistically relevant conclusions regarding SN Ia properties; which may further lead to a better understanding of dark energy and its properties.

In this context, the LSST team had hosted a competition on Kaggle\footnote{https://www.kaggle.com/c/PLAsTiCC-2018}, called Photometric LSST Astronomical Time-Series Classification Challenge (PLAsTiCC;  \citet{plasticc_2018}). The challenge was to correctly classify the test dataset into 15 astronomical classes whilst using a small training set.

Early deep learning approaches (\citet{hinners2018machine}; \citet{charnock2017deep}; \citet{revsbech2018staccato}) to photometric classification worked well on a test set representative of the training data but had poor performance when the test set was unrepresentative. However, later works (\citet{muthukrishna2019rapid}; \citet{pasquet2019pelican}; \citet{moller2020supernnova};  \citet{boone2019avocado};) were successful in this regard. In this text, we describe our approach to the PLAsTiCC problem using deep learning.

Our text is structured in the following way: We talk about our dataset in Section~\ref{sec:data}, and about our preprocessing and deep learning techniques in Section~\ref{sec:methods}. Finally, in Section~\ref{sec:results} we talk about our results obtained using this model and conclude in Section~\ref{sec:discussion}.

\section{Data}
\label{sec:data}

\begin{figure}
\label{fig:passbands}
	\includegraphics[width=\columnwidth]{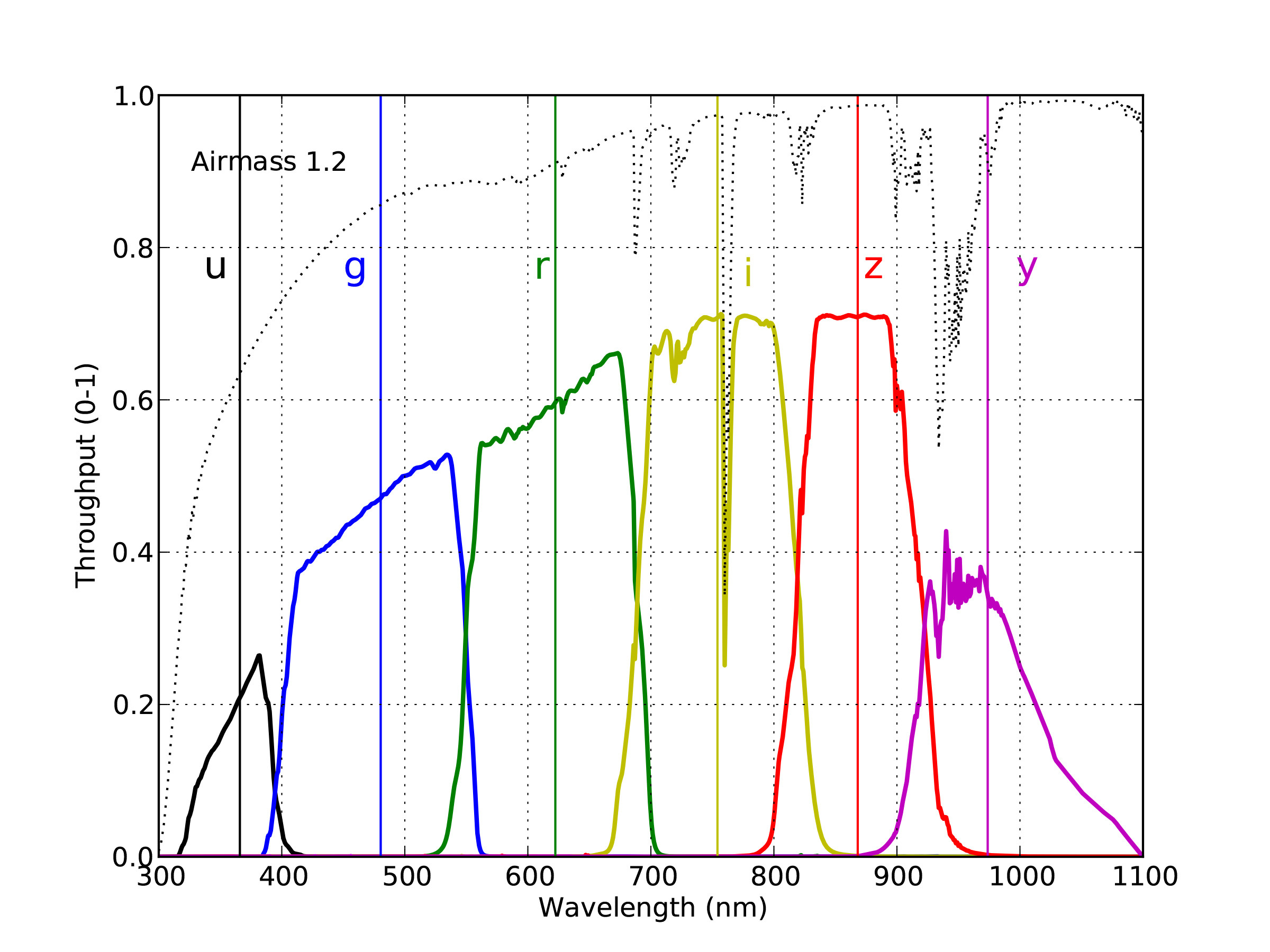}
    \caption{The LSST bandpasses. The vertical axis shows the total throughput. The computation includes the atmospheric transmission (assuming an airmass of 1.2, dotted line), optics, and the detector sensitivity.\newline Source: \citet{Ivezi_2019}}
    \label{fig:example_figure}
\end{figure}
\subsection{Overview}
For the PLAsTiCC Challenge, we will be dealing with univariate time series data. \emph{Time series data} is a series of observations of a variable of interest (in this case, brightness/intensity of light/flux) collected at multiple periods. In astronomy, such a series is called a \emph{light curve}; a graph of light intensity of an astronomical object, as a function of time. \emph{Spectroscopy} is the measurement of flux emitted across all wavelengths by an object. Spectroscopic light curves are the ultimate tool for classification as they result in the most reliable predictions. However, spectroscopic observations come at the cost of tremendous telescope time, and it is not possible to obtain spectroscopic data for every object observed (LSST expects an average of $10\times10^6$ alerts per night)\citep{collaboration2009lsst}. The goal is to narrow down the number of objects of interest for spectroscopic follow-up so that resources can be allocated on a priority basis. Photometry is the measurement of light through optical filters  ('passbands') that only permit a specific range of wavelengths to pass through. The LSST will make use of six passbands called \emph{u, g, r, i, z and y} as shown in Figure~\ref{fig:passbands} to measure the brightness of each object. The flux of light in each passband, measured as a function of time, is a light curve.

The PLAsTiCC dataset is a simulation of light curves under realistic observing conditions, as expected to be seen by LSST. The PLAsTiCC dataset consists of 3,500,734 light curves (7,846 in the training set and 3,492,888 in the test set) of various transient as well as variable objects from a mixture of different astrophysical classes, the full list of which is available in Table~\ref{tab:classes_table}.  PLAsTiCC takes the difficulty up a notch from SNPhotCC's supernova classification as it deals with this vast diversity of astrophysical classes.

\begin{table*}
	\caption{Summary of all the different object types and their counts in our dataset. \citep{Kessler_2019}}
	\label{tab:classes_table}
	\begin{tabular}{clcc} % four columns, alignment for each
		\hline
		\textbf{Class Number} & \textbf{Class Name} & \textbf{Number of Training Samples} & \textbf{Number of Test Samples}\\
		\hline
		90 &  Type Ia SN  & 2,313 & 1,659,831\\
		67 &  Peculiar Type Ia SN: 91bg-like  & 208 & 40,193\\
		52 &  Peculiar Type Ia SN: SNIax  & 183 & 63,664\\
		42 &  Type II SN  & 1,193 & 1,000,150\\
		62 &  Type Ibc SN  & 484 & 175,094\\
		95 &  Superluminous SN (Magnetar)  & 175 & 35,782\\
		15 &  Tidal disruption event  & 495 & 13,555\\
		64 &  Kilonova  & 102 & 133\\
		88 &  Active galactic nuclei  & 370 & 101,424\\
		92 &  RR Lyrae  & 239 & 197,155\\
		65 &  M-dwarf stellar flare  & 981 & 93,494\\
		16 &  Eclipsing binary stars  & 924 & 96,472\\
		53 &  Mira variables  & 30 & 1,453\\
		6 &  Microlens from single lens  & 151 & 1,303\\
		\hline
		991 &  Microlens from binary lens  & 0 & 533\\
		992 &  Intermediate luminous optical transient  & 0 & 1,702\\
		993 &  Calcium-rich transient  & 0 & 9,680\\
		994 &  Pair instability SN  & 0 & 1,1172\\
		\hline
	\end{tabular}
\end{table*}

\subsection{The PLAsTiCC Dataset in detail}
The PLAsTiCC data was simulated using 18 models developed by the members of the astronomy community at the behest of the PLAsTiCC team, 14 of which are based on enough observations to be represented in the training set (these form the 14 classes). Rest of the four were combined to form the 15th class found in the test set, as these objects have not been observed enough, or have never been observed but predicted to exist. This is the novel and truly challenging aspect of this competition - building classifiers that can identify unknown objects not present in the training set and correctly classify them into an unknown class. Using these models and three years of LSST observations, over 100 million transient and variable sources were simulated, of which 3.5 million satisfied the detection criteria explained in \citet{Kessler_2019}. These 3.5 million ugrizy light curves and the corresponding 453 million observations together form the PLAsTiCC dataset\citep{unblinded_plasticc}.  

The simulations assume that galactic objects have redshifts set to zero. For extragalactic objects, the simulation includes a model of a follow-up survey as described in \citet{Kessler_2019}. With this follow-up survey, 3.6\% of the extragalactic objects have spectroscopic redshifts for their hosts. Meanwhile, extragalactic objects without spectroscopic redshifts were assigned photometric redshifts obtained using a model described in (Kessler et al. 2019) and were accompanied by their uncertainties. The training set consisted of 7,848 objects with spectroscopically confirmed types, representing a mere 0.2\% of the total data (3,500,374 objects). The training set objects typically have a lower redshift than test set objects as spectra are necessary for type confirmation of each object, and it is easier to obtain spectroscopic data for nearby objects. Thus, the training set is biased towards brighter, closer objects.  

Another difference between objects arises due to the two distinct regions being monitored. LSST will probe two distinct regions called Deep Drilling Field (DDF) and Wide-Fast-Deep(WFD). DDF will cover 50 deg$^2$ and DDF observations are effectively around 1.5 mag deeper and 2.5 times more frequent than the WFD observations. Thus, objects in the DDF patches have well determined light curves with only small errors in flux. On the other hand, WFD observations will cover almost half of the sky, which will be observed less frequently, increasing uncertainties in the light-curves from objects belonging to this region. Our dataset contains a mix of DDF and WFD objects. While the relative areas of the DDF compared to the WFD for LSST will be 1/400, roughly 1\% of the simulated PLAsTiCC data are from the DDF subset \citep{plasticc_2018}.

The distribution of class sizes is wide, spanning from $ \sim10^2$ for the Kilonova class to $ \sim10^6$ for a few supernova types, which is quite apparent from  Table~\ref{tab:classes_table}. Many of the light curves are truncated because any given sky location is not visible (at night) from the LSST site for several months of the year. The data consists of irregular time series as it is not sampled at regular time intervals and different passbands are taken at different times, sometimes several days apart. The train: test split ratio is nearly 1:445, a highly skewed dataset with a significant variance in the number of examples of each class present in the training set. Thus our training set is highly imbalanced and non-representative, meaning that the distribution of objects in various classes is not uniform across the training set and test set. This imbalance is a massive departure from most deep learning problems, which tend to have representative datasets and extensive training sets.

The PLAsTiCC training set consists of two comma-separated values (CSV) files - a training set file which consists of light curves of all objects and a header file which consists of summary (astronomical) information available for each object. The PLAsTiCC test set is also similarly divided, with 11 light-curve files and one metadata file amounting to over 18 gigabytes of data.

\subsection{MetaData}
The header file lists each source in the data indexed by a unique identifier "object\_id", that is an integer. Each row of the table lists the properties of the source as follows:
\begin{enumerate}
    \item object\_id (\emph{int32}): Unique Object Identifier.
    \item ra (\emph{float32}): right ascension, sky coordinate: longitude, given in degrees.
    \item decl (\emph{float32}): declination, sky coordinate: latitude, given in degrees. 
    \item gal\_l (\emph{float32}): Galactic longitude, given in degrees. 
    \item gal\_b (\emph{float32}): Galactic latitude, given in degrees.
    \item ddf (\emph{bool}): A Boolean flag to identify the object as coming from the DDF survey area (with value ddf = 1 for the DDF). =
    \item hostgal\_specz (\emph{float32}): The spectroscopic redshift of the source. This is an extremely accurate measure of redshift, provided for the training set and a small fraction of the test set.
    \item hostgal\_photoz (\emph{float32}): The photometric redshift of the host galaxy of the astronomical source. A substitute for hostgal\_specz, this is more error prone.
    \item hostgal\_photoz\_err  (\emph{float32}): The uncertainty on the hostgal\_photoz based on LSST survey projections.
    \item distmod (\emph{float32}): The distance (modulus) calculated from the hostgal\_photoz redshift.
    \item MWEBV = MW E(B-V) (\emph{float32}): This 'extinction' of light is a property of the Milky Way (MW) dust along the line of sight to the astronomical source, and is a function of the sky coordinates of the source ra, decl.
    \item target (\emph{int8}): The class of the astronomical source. This is provided only for the training data.
\end{enumerate}

\subsection{Light Curve Data}
Each row of the light-curve table corresponds to an observation of the source at a particular time and passband. This includes the following information:
\begin{enumerate}
    \item object id (\emph{int32}): Same as in the metadata table, given as numbers
    \item mjd (\emph{float64}): The time of the observation, measured in Modified Julian Date (MJD), given in days. 
    \item passband (\emph{int8}): The specific LSST passband integer in which it was viewed, such that u, g, r, i, z, y = 0, 1, 2, 3, 4, 5.
    \item flux (\emph{float32}): the measured flux (brightness) in the passband of observation as listed in the passband column. The flux is corrected for MWEBV, but the error increases for larger MWEBV values. Note that the units for both flux and flux err are arbitrary.
    \item flux\_err (\emph{float32}): the uncertainty on the measurement of the flux listed above, given as float32 number.
    \item detected (\emph{bool}): This is given as a Boolean flag. The value is 1 if the object's brightness is significantly different at the 3$\sigma$ level relative to the reference template, and 0 otherwise.
\end{enumerate}

Each night, it is expected that LSST will produce 15 Terabytes of imaging data, and up to $10^7$ transient detections to sift through and find exciting candidates to analyse and to target for spectroscopic observations. This calls for a need to develop early epoch classification based on limited observations so that spectroscopic observations can be scheduled on exciting subsets. Keeping this in mind, we have tried to minimise the use of computational resources and optimise computation time to as great an extent as possible. We have dropped features like sky coordinates, galactic coordinates and the "detected" Boolean flag to reduce the dimensionality of our feature space. This also made it possible for us to handle such a large dataset with limited resources at our disposal.

\section{Methods}
\label{sec:methods}

\subsection{Overview}
In this section, we have described the techniques used by us for the photometric classification of light curves. First, we preprocess the given data and extract features from it as described in Section~\ref{sub:preprocess}. We have then described evaluation metrics considered in ~\ref{sub:metrics} and loss functions considered in ~\ref{sub:losses}.  Our deep learning model consists of a dense-based submodel:  \textbf{2DSubM} and a GRU-based submodel: \textbf{3DSubM}, each named after the shape of their input, respectively. This has been described in Section~\ref{sub:architecture}, with the hyperparameters in Section~\ref{sub:hyperparameters}. We then describe our Ensemble technique in Section~\ref{sub:ensemble}, which resulted in a significant boost to our model. We finally talk about the augmentation techniques we tried in Section~\ref{sub:augmentation}, and why they were not useful.

\subsection{Preprocessing}
\label{sub:preprocess}
Our input data consists of 2 files each for training and testing - light curve data and metadata, as described in Section~\ref{sec:data}. We preprocess the available data in two separate ways based on the input requirements for the two types of submodels. Our preprocessing consists of the following steps:  
\subsubsection{2DSubM Data}
\label{subsub:2DSubM}
\begin{enumerate}
    \item Adding noise to the flux in the form of a normal distribution having mean zero and standard deviation equal to the $\frac{2}{3}$ times the error in flux (i.e. flux\_err).
    \item Scaling down the flux by multiplying every flux value by a quantity we define as:
        \begin{equation}
        \label{eq:scaler}
            flux\_scaler =\frac{1}{log_2(flux\_max - flux\_min)}
        \end{equation}
    where $flux\_max$ and $flux\_min$ are the maximum and minimum flux present in the training set, respectively.
    \item Aggregating a total of 156 features from the light curve data, based on the features provided by Siddhartha (meaninglesslives) on Kaggle\footnote{\url{https://www.kaggle.com/meaninglesslives/simple-neural-net-for-time-series-classification\#Extracting-Features-from-train-set}}.
    \item Merging the above-aggregated features with the metadata, and dropping the features that are not required.
    \item Replacing all NaNs with zeros.
    \item Power Transforming the 2D data using the method as described by Yeo \& Johnson \citep{power_transformer}.
\end{enumerate}

\subsubsection{3DSubM Data}
\label{subsub:3DSubM}
\begin{enumerate}
    \item Adding noise to the flux in the form of a normal distribution having mean zero and standard deviation equal to the $\frac{2}{3}$ times the error in flux (i.e flux\_err).
    \item Scaling down the flux by multiplying every flux value by the flux\_scaler as defined in (Eq.~\ref{eq:scaler})
    \item Calculating the difference in dates (MJDs) between observations and using that as a feature instead of the raw MJD value.
    \item Grouping all observations occurring on the same night(i.e. within 8 hours of each other) together on the basis of their passbands. This gives us a table per object, consisting of various dates and the flux for each passband on that day. In case of multiple observations in a passband on the same night, the entry with the lower error is chosen. In case of no observations for any passband, the date is skipped.
    \item Filling all missing flux entries for a particular passband using linear interpolation(with respect to time) using the preceding and succeeding entries. 
    \item Replacing all NaNs with zeros.
    \item Padding each day's observations with zeroes. Since different objects have a different number of observations and thus different length, we pad them in order to make all sequences equal to the maximum length in the training set (162 in our case).
\end{enumerate}

\subsection{Evaluation Metrics}
\label{sub:metrics}
\subsubsection{Accuracy:}
\emph{Accuracy} is given by the number of correctly classified examples divided by the total number of classified examples \citep{100pagemlbook}. In terms of the confusion matrix, it is given by:
\begin{equation}
    \text{Accuracy} = \frac{TP + TN}{TP + TN + FP + FN}
\end{equation}
where,\newline TP: True Positive\newline TN: True Negative\newline FP: False positive\newline FN: False Negative
\newline

Accuracy is a useful metric when errors in predicting all classes are equally important.
In our model, we have assumed equal weights for all classes. 

\subsubsection{Precision:}
\emph{Precision} is the ratio of correct positive predictions to the overall number of positive predictions \citep{100pagemlbook}. Intuitively, the precision is the ability of the classifier not to label as positive a sample that is negative. It is given by:
\begin{equation}
    \text{Precision} = \frac{TP}{TP + FP}
\end{equation}

\subsubsection{Recall:}
\emph{Recall} is the ratio of correct positive predictions to the overall number of positive examples in the dataset \citep{100pagemlbook}. Intuitively, the recall is the ability of the classifier to find all the positive samples. It is given by:
\begin{equation}
    \text{Recall} = \frac{TP}{TP + FN}
\end{equation}

In multiclass classification, precision/recall are evaluated for a particular class. This is done by considering all examples of the selected class as positives and all examples of the remaining classes as negatives. On Python, scikit-learn provides various ways to evaluate precision scores by using the 'average' parameter. We will be using 'macro' precision/recall for our final model evaluation along with accuracy. Macro precision/recall calculates metrics for each label and finds their unweighted mean. This does not take label imbalance into account.

To get an understanding of the performance, we performed 5-fold cross-validation \citep{brownlee2018kfold} on the training set, and then selected the best model based on the average accuracy. Then, a random search of the hyperparameter space was performed, as described in Section~\ref{sub:hyperparameters}. After obtaining the best model, it was retrained from scratch on the whole training set.

\subsection{Loss Functions}
\label{sub:losses}
We experimented with three different loss functions:
\subsubsection{Categorical Cross Entropy}
Categorical Cross entropy is the loss function used commonly when dealing with multi-class classification that requires output to be a probability value between 0 and 1. Cross entropy loss increases as the predicted probability diverges from the actual target/label. 

\subsubsection{Focal Loss}
Focal Loss is a loss function generalising binary and multiclass cross-entropy loss that penalises hard-to-classify examples. It reshapes the loss function to down-weight easy examples and thus focuses training on hard negatives by adding a modulating factor $(1 - p_{t}) ^{\gamma} $ to the cross-entropy loss, with tunable focusing parameter $\gamma \geq 0$.

\begin{equation}
FL (p_t) = - \alpha_t (1-p_t)^\gamma log(p_t)
\end{equation}
Focal Loss is mainly used in classification problems dealing with class imbalance, with a sparse set of hard examples \citep{focal_loss}.

\subsubsection{LogLoss}
LogLoss is defined as the negative of log-likelihood and is generally used for probabilistic classification rather than deterministic classification. Valid PLAsTiCC entries had to include probability scores for each class and each astronomical source, with the aim of minimising the PLAsTiCC metric score which is defined as a weighted log-loss metric. We did not experience an improvement in performance with LogLoss and thus discarded it.

\subsection{Model Architecture}
\label{sub:architecture}
We have used two types of submodels in our approach. The first is based on deeply connected dense layers, which we call \emph{2DSubM}, while the second type, \emph{3DSubM} is based on a combination of Gated Recurrent Unit (GRU; \citet{cho2014learning}) layers and dense layers. The text in this section will describe the top two models(of each type) arrived at after hyperparameter tuning, whose details have been described in Section~\ref{sub:hyperparameters}. The two different submodels are trained separately and then later combined using a method called stacking ensemble, in Section~\ref{sub:ensemble}.

\begin{figure}
    \centering
	\includegraphics[width=200pt]{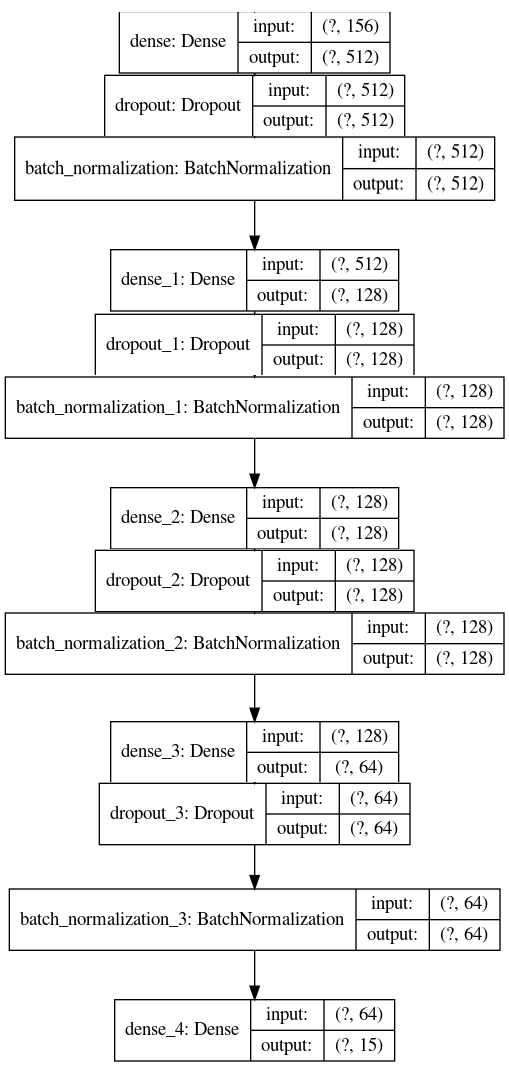}
    \caption{A schematic representation of our best 2DSubM model architecture}
    \label{fig:best_dense}
\end{figure}

\begin{figure}
    \centering
	\includegraphics[width=200pt]{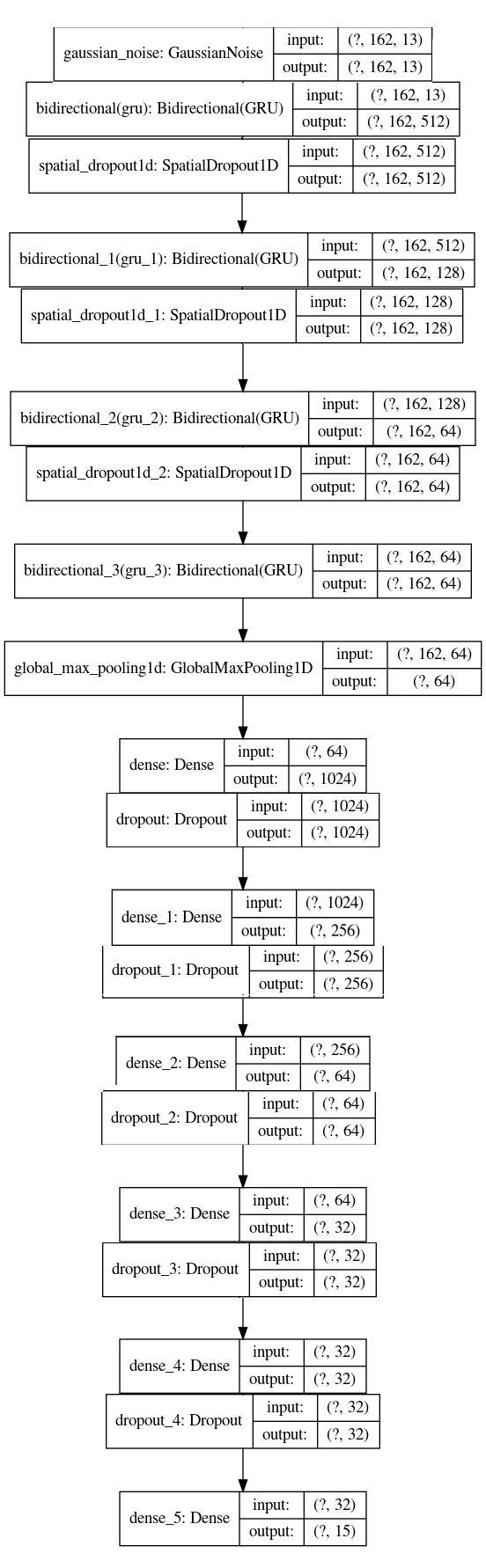}
    \caption{A schematic representation of our best 3DSubM model architecture}
    \label{fig:best_gru}
\end{figure}

\subsubsection{2DSubM Architecture}
The core of this submodel network consists of fully connected dense layers, taking a two-dimensional array as an input and classification probability as an output. Our best performing network consists of four dense layers consisting of 512, 128, 128 and 64 units, respectively.  All of the dense layers are activated using the tanh function, and a dropout of 0.1 and batch normalisation is applied after every dense layer. These help with regularisation and improve training. The output layer is a softmax layer of 15 units. A schematic representation of the 2DSubM model architecture is available in Fig.~\ref{fig:best_dense}.

The \emph{2DSubM} network takes the preprocessed \emph{2DSubM} Data described in Section~\ref{subsub:2DSubM} as its input, which consists of metadata and aggregated features. This network was trained using Adam \citep{kingma2014adam} as the optimiser (learning rate = 0.01), and Categorical Cross Entropy as the loss function. The average 5-fold cross-validation accuracy of the \emph{2DSubM} architecture was found to be 0.773066.

\subsubsection{3DSubM Architecture}
This submodel network consists of a series of layers of Bidirectional GRUs and Dense layers. This network takes a three-dimensional array as an input, of shape (samples, timesteps, features). A GaussianNoise layer is applied at the start, which adds some noise to the inputs as a form of regularisation. This is followed by a series of Bidirectional GRUs, with the output of the last GRU layer being downsampled to a two-dimensional array by the GlobalMaxPooling1D layer. These two-dimensional arrays are then fed to a series of dense layers, which finally returns an output layer consisting of the classification probability. Our best performing network consisted of a GaussianNoise layer with a standard deviation of 0.5, followed by four Bidirectional GRU layers consisting of 256, 64, 32 and 32 units, respectively. All of the GRU layers are activated using the tanh function, and a spatial dropout of 0.1 is applied after every GRU layer, for regularisation. This is then followed by 4 densely connected layers consisting of 1024, 256, 64 and 32 units, respectively. All the dense layers are also tanh activated, and followed by a dropout of 0.1. Finally, the output layer is a softmax layer of 15 units. A schematic representation of the 3DSubM model architecture is available in Fig.~\ref{fig:best_gru}.

This network takes the preprocessed \emph{3DSubM} Data described in Section~\ref{subsub:3DSubM} as its input, which consists of the light curve data. This network was also trained using Adam as the optimiser (but with the default learning rate = 0.001 this time), while Focal Loss was chosen as the loss function. The average 5-fold cross validation accuracy obtained was 0.764018.

\subsection{Hyperparameter Tuning}
\label{sub:hyperparameters}

\begin{table}
	\centering
	\caption{List of Hyperparameters tested with Keras-Tuner}
	\label{tab:hyperparamter}
    \begin{tabular}{cc}
    \hline
    \textbf{Hyperparameters} & \textbf{Tested Settings}                      \\
    \hline
    Number of Dense Layers & {[}1,2,3,4,5,6,7{]}                             \\
    Number of GRU Layers   & {[}0,1,2,3,4,5,6{]}                             \\
    Number of Dense Units  & {[}128,256,512,1024{]}                          \\
    Number of GRU Units    & {[}64,128,256{]}                                \\
    Activation Function    & {[}tanh,relu{]}                                 \\
    Loss Function          & {[}categorical\_crossentropy, focal, logloss{]} \\
    Learning Rate          & {[}0.1, 0.01, 0.001, 0.0001{]}                  \\
    Dropout                & {[}0, 0.01, 0.1, 0.2, 0.5{]}                    \\
    SpatialDropout1D       & {[}0, 0.01, 0.1, 0.2, 0.5{]}                    \\
    GaussianNoise          & {[}0, 0.25, 0.5, 0.75, 1, 2, 5{]}               \\
    Bidirectionality       & {[}True, False{]}                               \\
    BatchNormalization     & {[}True, False{]}                               \\   
    \hline
	\end{tabular}
\end{table}

We used Keras-Tuner for automating our hyperparameter tuning. We performed an automated random search on the hyperparameter for various parameters, of which we chose the best parameters as described in Section~\ref{sub:architecture}. The full list of the hyperparameters is in Table~\ref{tab:hyperparamter}.

\begin{figure*}
    \centering
	\includegraphics[width=480pt]{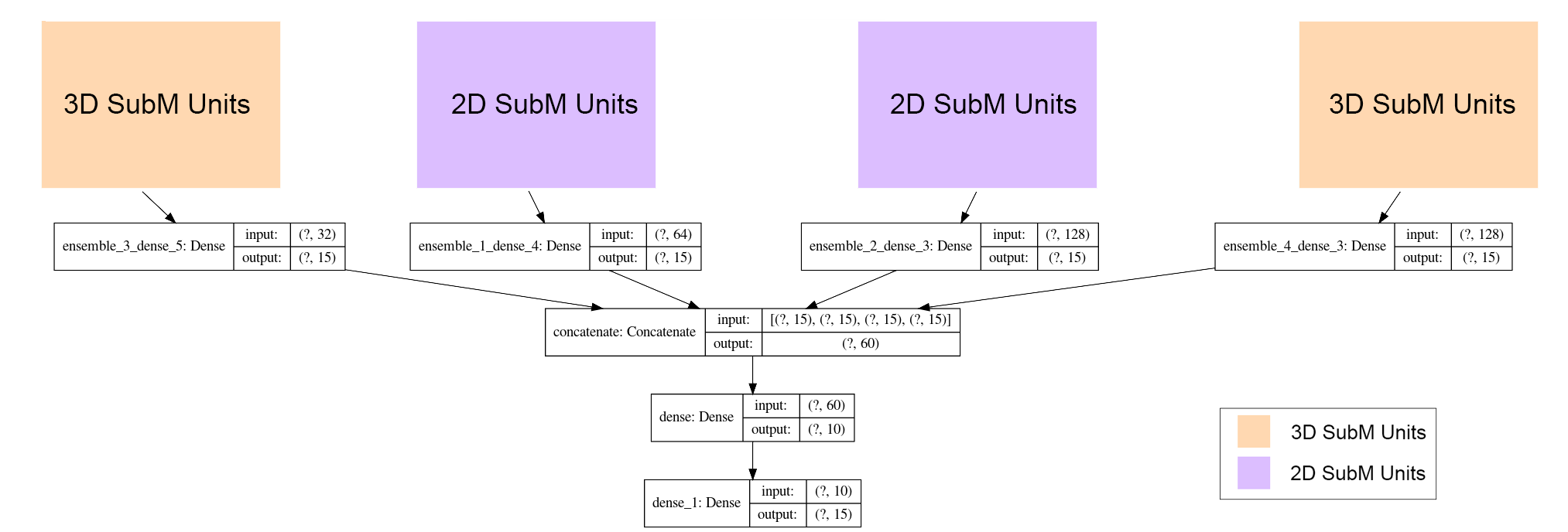}
    \caption{A schematic representation of our stacking ensemble}
    \label{fig:ensemble}
\end{figure*}

\subsection{Stacking Ensemble}
\label{sub:ensemble}
Instead of using a single model for prediction, we combined our best performing models by a technique called Stacking Ensemble \citep{Wolpert1992}. The top two performing models each of type \emph{2DSubM} and \emph{3DSubM} (thus a total of 4 submodels) were stacked, and their outputs were combined to form the inputs of a meta learner neural network. A dense layer of 10 units followed this, and finally, a softmax layer of 15 units was used as the output. The weights of the original submodels were set to be untrainable, allowing us to modify only the weights of the last two dense layers. This ensemble was then finally trained on our original validation set, with the Adam optimiser and Categorical Cross Entropy loss function for a total of 50 epochs. A schematic representation is available in Fig~\ref{fig:ensemble}. This was then saved as the final model to be used on our test data. The presence of the two types of submodels gives our architecture more diversity.

Our stacking ensemble can thus be described as a neural network which combines our pre-trained neural networks in appropriate proportions to get the best result.

\subsection{Data Augmentation}
\label{sub:augmentation}
Data Augmentation is the technique of increasing training set examples or populating the data available for training the model without actually collecting any new raw data. This technique is especially useful in cases when the train data to test data ratio is skewed, with very few training examples available as well as cases of imbalanced data, when the distribution across classes is not uniform. It is challenging to devise data augmentation techniques for time series data, especially irregular time series, and this has been covered to a far lesser extent in the literature compared to data augmentation for images.

\subsubsection{Gaussian Process}
A Gaussian Process(GP) is a probability distribution over possible functions.  This method of augmentation was implemented in {\fontfamily{qcr}\selectfont Avocado}, the top-scoring model of PLAsTiCC \citep{boone2019avocado} and in {\fontfamily{qcr}\selectfont STACCATO} \citep{revsbech2018staccato}. It involves fitting GPs to the observations of each object in the training set. Then, synthetic light curves are produced for each object by downsampling and dropping random points from the GPs. This is repeated multiple times, and because of the stochastic nature, each iteration downsamples the GP differently. This gives us a set of new synthetic light curves which are consistent with the observations of the original object. By degrading well-sampled light curves, and adding noise, it is thus possible to generate light curves with similar features as the test set and ensures that the augmented training set is more representative of the test set. Further, a large number of features can then be extracted from these GP-fitted curves, to be fed to the machine learning model.

The chief difficulty that we faced in implementing GP regressions was the lack of computational resources. Gaussian Process regressions are computationally expensive. K. Boone required about 100 core hours of computing time after using an Intel Xeon CPU, which was not an option for us.

\subsubsection{Redshift Modifications using Normal Distribution}
This method was implemented in the second-highest scoring model of PLAsTiCC\footnote{\url{https://www.kaggle.com/c/PLAsTiCC-2018/discussion/75059\#446148}}. The preprocessing part involved removing redshift dependency from all time and wavelength related features. This was done by assuming a linear relation on redshift on these features and then dividing them by (1 + hostgal\_photoz).  Then, up to 30\% of the observations  were dropped randomly. Redshift was then modified using a normal distribution with the standard deviation:
\begin{equation}
    \sigma_{\text{redshift}}  = \text{hostgal\_photoz\_err} \times \frac{2}{3} 
\end{equation}
 All time and wavelength related features were modified accordingly. Finally, flux was modified using a normal distribution with standard deviation:
 \begin{equation}
    \sigma_{\text{flux}} = \text{flux\_err} \times \frac{2}{3} 
\end{equation}

When we implemented this augmentation, our model did not show any improvement. We also tried replicating the results achieved by the authors, but our accuracies did not match their results. We later came to know that they had implemented neural network models in conjunction with LightGBM models in an ensemble. Our implementation of this technique yielded accuracies in the 35-40\% range and did not improve our performance. Hence, we discarded these data augmentation techniques, choosing instead to work with data modifications and scaling/normalising of features.

\subsection{Data Availability}
The training and test datasets used by us are available on PLAsTiCC's page on Kaggle here: \url{https://www.kaggle.com/c/PLAsTiCC-2018/data}. The unblinded test metadata to calculate the evaluation metrics was obtained from here: \url{https://zenodo.org/record/2539456}. Our pre-trained models and Python code can be found here: \url{https://github.com/AKnightWing/Astronomical-Classification-PLASTICC}.

\section{Results}
\label{sec:results}
We ran our final stacked model, as mentioned in Section~\ref{sub:ensemble}, on the test set. The scores achieved by our model on the metrics explained in Section~\ref{sub:metrics} are listed in Table~\ref{tab:results}. Preprocessing the entire test set (11 files, amounting to 18.43 gigabytes) took nearly 15 hours to complete. We used Kaggle Notebooks to train and evaluate our model. Training on an NVIDIA TESLA P100 GPU provided by Kaggle, took around 2 hours, while evaluation took just under 3 hours.

\begin{figure*}
	\includegraphics[width=480pt]{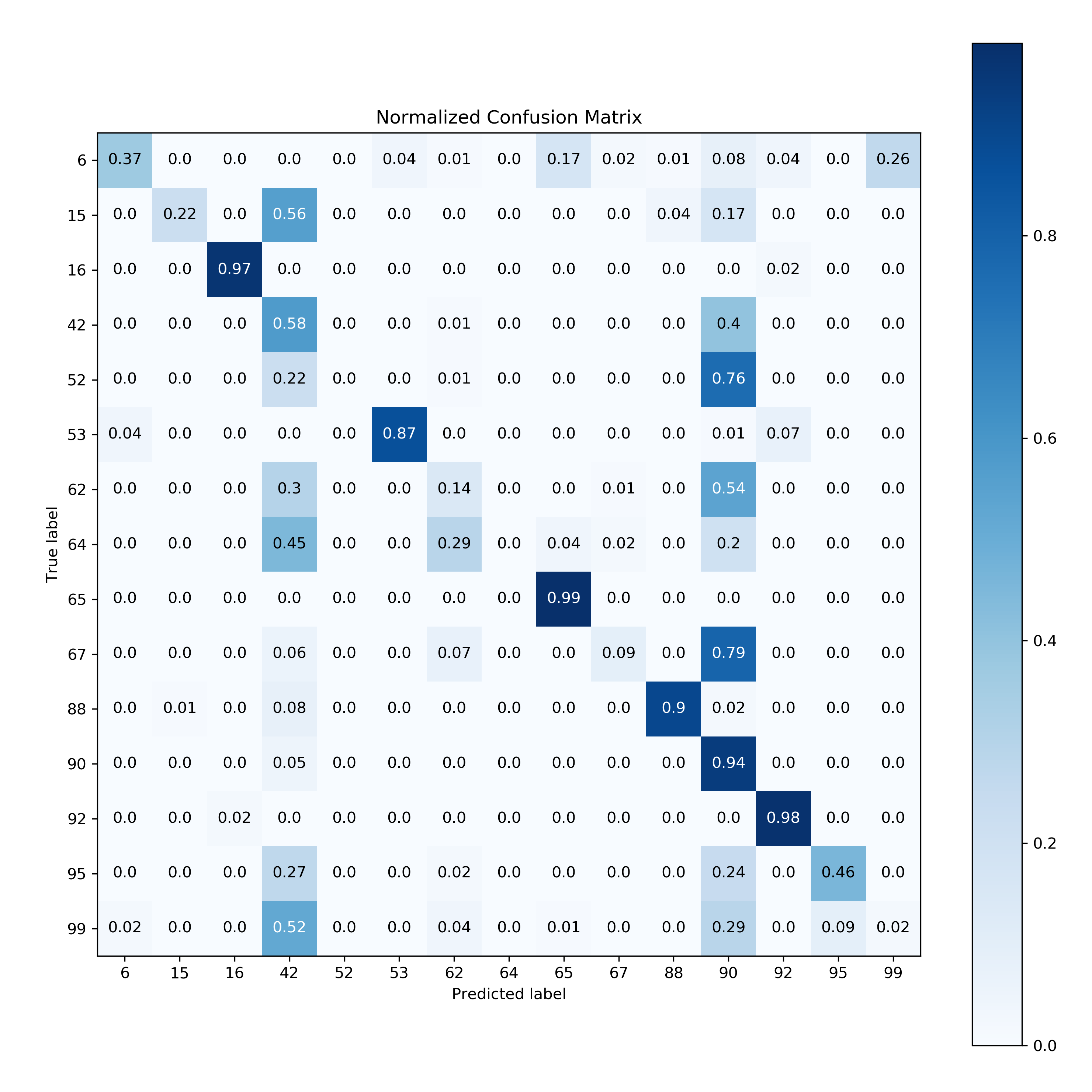}
    \caption{Confusion Matrix of our final stacked model on all of the test data: All the counts have been normalised to unity.}
    \label{fig:confusion_matrix}
\end{figure*}

Figure~\ref{fig:confusion_matrix} shows the confusion matrix generated post evaluation. Our model classified classes 16, 65 and 92 exceptionally well with accuracies 97\%, 99\% and 98\% respectively. These classes are eclipsing binary stars, M-dwarf stellar flares and RR Lyrae respectively which have their characteristic light curves and our model successfully learned to classify them using this information. The next best-classified objects were Active Galactic Nuclei (AGN) and Type Ia Supernovae. Our model performed the worst when classifying Peculiar Type Ia SN (both 91bg-like and SNIax) and Type Ibc SN, all three of which were predicted to be Type Ia SN most frequently. Kilonova events were most frequently misclassified as Type II SN. We will discuss the possible reasons behind these in Section~\ref{sec:discussion}.

\begin{table}
	\centering
	\caption{Table Summarising the Metrics Scores produced by our final stacked model on all of the test data.}
	\label{tab:results}
	\begin{tabular}{cc}
		\hline
		\textbf{Metric} & \textbf{Score}\\
		\hline
		Accuracy & 76.243 \%\\
		Precision & 0.62266\\
		Recall & 0.50290\\
		\hline
	\end{tabular}
\end{table}

\section{Discussion and Conclusion}
\label{sec:discussion}

Our model performed poorly on some classes, as mentioned in Section~\ref{sec:results}. Kilonova events were underrepresented in both the training set as well as the test set with a total of only 231 objects in the entire dataset. The light curves for Type Ib and Type Ic supernovae have been found to be nearly identical to Type Ia supernovae in some cases. Also, SN Ib do not differ much from SN Ia when comparing the absolute magnitude at peak brightness \citep{Tsvetkov}. This would explain why the model poorly classifies Type Ibc SN, labelling them as Type Ia SN instead. Poor classification of these classes can be attributed to underrepresentation of that class in the training set, rare instances of occurrence in test data (specially Kilonova) and similarity with the major class of the test set. 

Our model performed poorly on some classes, as mentioned in Section~\ref{sec:results}. Kilonova events were underrepresented in both the training set as well as the test set with a total of only 235 objects in the entire dataset. The light curves for Type Ib and Type Ic supernovae have been found to be nearly identical to Type Ia supernovae in some cases. Also, SN Ib do not differ much from SN Ia when comparing the absolute magnitude at peak brightness \citep{Tsvetkov}. This would explain why the model poorly classifies Type Ibc SN, labelling them as Type Ia SN instead. Poor classification of these classes can be attributed to the underrepresentation of that class in the training set, rare instances of occurrence in test data (especially Kilonova) and similarity with the major class of the test set. 

These drawbacks can be overcome by implementing computationally intensive data augmentation techniques like GP fits which have been shown to significantly improve results by generating a more representative training set. We could also calculate and include more relevant features obtained from the literature pertaining to astronomical objects. These may help to distinguish between similar classes. Another way to improve model performance would be to build a model that predicts hostgal\_specz as a meta encoder. Working with minimal computational resources, our model could achieve up to 76\% accuracy. With more computational resources, we could implement different preprocessing strategies and include more models in the final stacked ensemble to boost performance. 

Another interest of ours would be to compare the performance of this model on other light curve datasets. Light curve data from the Sloan Digital Sky Survey(SDSS; \citet{Frieman_2007}) and the Supernova Photometric Classification Challenge (SNPhotCC; \citet{kessler2010supernova}) are suitable options. The new data might provide further insights to make our current model more robust.

\section*{Acknowledgements}

We thank Dr. Parthiban Srinivasan for his guidance, support and invaluable inputs. We thank the LSST Team for coming up with this innovative challenge. We thank the PLAsTiCC organisers for the simulated data set and for organising this challenge. We thank Kaggle for providing a platform to host PLAsTiCC. We also thank the Kaggle users over at the PLAsTiCC Discussion Boards for the helpful comments and ideas shared.

\emph{Software:} Keras \citep{chollet2015keras}, Keras-Tuner \citep{omalley2019kerastuner}, NumPy(\citet{numpy}; \citet{numpy2}), Jupyter \citep{jupyter}, Matplotlib \citep{matplotlib}, scikit-learn \citep{scikit-learn}, SciPy \citep{scipy}, Pandas(\citet{pandas}; \citet{pandas2}), TensorFlow \citep{tensorflow} and Python3. \citep{python3}

%%%%%%%%%%%%%%%%%%%%%%%%%%%%%%%%%%%%%%%%%%%%%%%%%%



\bibliographystyle{mnras}
\bibliography{bibliography.bib} % if your bibtex file is called example.bib

%\begin{thebibliography}{99}
%\end{thebibliography}

% Don't change these lines
\bsp	% typesetting comment
\label{lastpage}
\end{document}